\documentclass[twocolumn,amsmath,aps]{revtex4}
\usepackage{graphicx}

\newcommand{\nl}{\nonumber \\}

\newcommand{\be}{\begin{equation}}
\newcommand{\ee}{\end{equation}}
\newcommand{\bea}{\begin{eqnarray}}
\newcommand{\eea}{\end{eqnarray}}

\newcommand{\Eq}[1]{Eq.\,(\ref{#1})}
\newcommand{\Eqs}[1]{Eqs.\,(\ref{#1})}

\newcommand{\la}{\langle}
\newcommand{\ra}{\rangle}
\newcommand{\dg}{\dagger}

\newcommand{\ti}{\tilde}
\newcommand{\mb}{\mbox}

\begin{document}
\draft

\title{Spontaneous Relaxation of a Charge Qubit under Electrical Measurement}

\author{ Xin-Qi Li$^{1,2}$, Ping Cui$^{2}$, and YiJing Yan$^{2}$}

\address{$^1$ Institute of Semiconductors,
         Chinese Academy of Sciences, P.O.~Box 912, Beijing 100083, China}
\address{$^2$ Department of Chemistry, Hong Kong University of Science and
         Technology, Kowloon, Hong Kong }
\date{\today}

\begin{abstract}
In this work we first derive a generalized conditional master
equation for quantum measurement by a mesoscopic detector, then
study the readout characteristics of qubit measurement where a
number of new features are found.
The work would in particular highlight the qubit spontaneous relaxation
effect induced by the measurement itself rather than
an external thermal bath. 
\end{abstract}
\maketitle


The recent renewed interest of measuring a two-state quantum
system (qubit) stems largely from the rapidly developing field of
quantum computing. A possible solid-state implementation of such
measurement is to measure a charge qubit by a mesoscopic detector
which, for instance, can be either a quantum-point-contact (QPC)
\cite{Gur97,Win97,Moz02,Gur03}, or a single-electron-transistor
(SET) \cite{Sch98}.

For a realistic setup of such measurement, the non-trivial
correlation between the detector and the qubit has been the focus
of recent theoretical studies. However, sometimes the treatment of
this correlation is incomplete. For instance, in a number of
publications on the qubit measurement by a QPC
\cite{Gur97,Gur03,Kor01a,Goa01a}, the energy transfer
between the detector and qubit has been ignored, which leads the
qubit to an incorrect statistical mixture under low measurement
voltage, as shown in our work \cite{Li04}.
Two recent publications considered the energy-exchange induced
inelastic effect on the detector power spectrum
by using, respectively, the real-time Green's function approach \cite{Shn02}
and the quantum jump technique \cite{Sta03},
where a number of controversial results were arrived
and cause further debate \cite{Ave04}.

In this paper, by generalizing the work of Gurvitz {\it et al.}
\cite{Gur97,Moz02,Gur03},
we present an alternative approach to study the inelastic effect in the
qubit measurement by a QPC.
Connections with the previous work will be established in a transparent way,
and new features will be illustrated in both the output current and
power spectrum.
In particular, we shall highlight the qubit {\it spontaneous relaxation} effect
induced by energy exchange with the measuring device, instead of coupling
to an {\it external} thermal bath as discussed in Ref.\ \onlinecite{Gur03}.


For the sake of generality, we first formally consider an arbitrary
quantum system measured by a QPC, described by
\begin{subequations} \label{H1}
\begin{eqnarray}
H   &=&  H_0+H'  ,
\\
H_0 &=&  H_{s}   +  \sum_k (\epsilon^L_k c^{\dg}_kc_k+
         \epsilon^R_k d^{\dg}_kd_k) ,
\\
H'  &=& \sum_{k,q} [T_{qk}\{|\psi_s\ra\la \psi_s|\}
         d^{\dg}_q c_k  + \mb{H.c.}] .
\end{eqnarray}
\end{subequations}
In this decomposition, the free part of the total Hamiltonian $H_0$
contains the Hamiltonians of the measured system $H_s$
and the QPC reservoirs (the last two terms).
The interaction Hamiltonian $H'$ describes electron tunneling through the QPC,
e.g., from state $|k\ra$ in the left reservoir to
state $|q\ra$ in the right reservoir,
with tunneling amplitude $T_{qk}\{|\psi_s\ra\la \psi_s|\}$
that is conditioned by the eigenstate $|\psi_s\ra$ of the observable.


Regarding the tunneling Hamiltonian $H'$ as perturbation,
on the basis of the second-order cummulant expansion we can derive
a formal equation for the reduced density matrix as \cite{Yan98}
\bea\label{ME-1}
\dot{\rho}(t) = -i {\cal L}\rho(t) - \int^{t}_{0}d\tau \la {\cal L}'(t){\cal G}(t,\tau)
                {\cal L}'(\tau){\cal G}^{\dg}(t,\tau) \ra \rho(t).
\eea
Here the Liouvillian superoperators are defined as
${\cal L}(\cdots)\equiv [H_s,(\cdots)]$,
${\cal L'}(\cdots)\equiv [H',(\cdots)]$, and
${\cal G}(t,\tau)(\cdots)\equiv G(t,\tau)(\cdots)G^{\dg}(t,\tau)$
with $G(t,\tau)$ the usual propagator (Green's function) associated with $H_s$.
The reduced density matrix $\rho(t)=\mb{Tr}_D[\rho_T(t)]$,
resulting from tracing out
all the detector degrees of freedom from the entire density matrix.
However, for quantum measurement where the specific readout information
is likely to be recorded, the average should be performed
over the unique class of states of the detector
we are trying to keep track of.

For the measurement setup under study, the relevant quantity of readout
is the transport current $i(t)$ in the detector,
or equivalently, the number of electrons that have tunnelled
through the detector, $n(t)=\int^{t}_{0} dt' i(t')$.
We therefore classify the Hilbert space of the detector as follows.
First, we define the subspace in the absence of electron tunneling
through the detector as ${\cal D}^{(0)}$, which is spanned by the product
of all many-particle states of the two isolated reservoirs, formally denoted
as ${\cal D}^{(0)}\equiv\mb{span}\{|\Psi_L\ra\otimes |\Psi_R\ra \}$.
Then, we introduce the tunneling operator
$f^{\dg}\sim f^{\dg}_{qk}=d_q^{\dg}c_k$,
and denote the Hilbert subspace corresponding to $n$-electrons
tunnelled from the left to the right reservoirs
as ${\cal D}^{(n)}=(f^{\dg})^n {\cal D}^{(0)}$, where $n=1,2,\cdots$.
The entire Hilbert space of the detector is ${\cal D}=\oplus_n{\cal D}^{(n)}$.

With the above classification of the detector states, the average over states
in ${\cal D}$ in \Eq{ME-1} is replaced with states in the subspace ${\cal D}^{(n)}$,
leading to a {\it conditional} master equation
\bea\label{ME-2}
\dot{\rho}^{(n)}(t) &=& -i {\cal L}\rho^{(n)}(t) - \int^{t}_{0}d\tau
      \mb{Tr}_{D^{(n)}} [{\cal L}'(t){\cal G}(t,\tau)   \nl
      & &  \times {\cal L}'(\tau) {\cal G}^{\dg}(t,\tau) \rho_T(t)] .
\eea
Here $\rho^{(n)}(t)=\mb{Tr}_{D^{(n)}}[\rho_T(t)]$,
which is the reduced density matrix of the measured system
{\it conditioned} by the number of electrons tunnelled through the detector
until time $t$.
Now we transform the Liouvillian operator product in \Eq{ME-2}
into the conventional Hilbert form:
\bea\label{L-H}
&&  {\cal L}'(t){\cal G}(t,\tau) {\cal L}'(\tau) {\cal G}^{\dg}(t,\tau) \rho_T(t)   \nl
&=& [ H'(t)G(t,\tau)H'(\tau)G^{\dg}(t,\tau)\rho_T(t)  \nl
& &     - G(t,\tau)H'(\tau)G^{\dg}(t,\tau)\rho_T(t)H'(t)] + \mb{H.c.} \nl
&\equiv& [I-II]+\mb{H.c.}
\eea

For the convenience of description, we rewrite the interaction Hamiltonian as
$H'(t)=QF(t)$.
Here we have assumed the tunneling amplitude $T_{kq}$ to be real and
independent of the reservoir-state ``$kq$" ,
and denoted it by $Q$ which depends on the state of the measured system.
The detector fluctuation is described by $F(t)\equiv f(t)+f^{\dg}(t)$,
with $f\equiv\sum_{kq}c^{\dg}_{k}d_q$ and $f^{\dg}\equiv\sum_{kq}d^{\dg}_{q}c_k$.
To proceed, two physical considerations are further involved as follows:
(i)
Instead of the conventional Born approximation for the entire density matrix
$\rho_T(t)\simeq\rho(t)\otimes\rho_D$,
we propose the ansatz $\rho_T(t)\simeq\sum_n\rho^{(n)}(t)\otimes\rho_D^{(n)}$,
where $\rho_D^{(n)}$ is the density operator of the detector reservoirs with
$n$-electrons tunnelled through the detector.
With the ansatz of the density operator, tracing over the subspace
${\cal D}^{(n)}$ yields
\begin{subequations}\label{n-ave}
\bea
\mb{Tr}_{D^{(n)}}[I]&=& \mb{Tr}_D [F(t)F(\tau)\rho_D^{(n)}]  \nl
       & & \times [QG(t,\tau)QG^{\dg}(t,\tau)\rho^{(n)}]  \\
\mb{Tr}_{D^{(n)}}[II]&=& \mb{Tr}_D [f^{\dg}(\tau)\rho_D^{(n-1)}f(t)] \nl
       & & \times [ G(t,\tau)QG^{\dg}(t,\tau)\rho^{(n-1)}Q ]   \nl
       & & + \mb{Tr}_D [f(\tau)\rho_D^{(n+1)}f^{\dg}(t)] \nl
       & & \times [ G(t,\tau)QG^{\dg}(t,\tau)\rho^{(n+1)}Q ] .
\eea
\end{subequations}
Here we have utilized the orthogonality between states in different subspaces,
which in fact leads to the term selection from the entire density operator $\rho_T$.
(ii)
Due to the closed nature of the detector circuit, the extra electrons
tunnelled into the right reservoir will flow back into the left reservoir
via the external circuit.
Also, the rapid relaxation processes in the reservoirs will quickly
bring the reservoirs
to the local thermal equilibrium state determined by the chemical potentials.
As a consequence, after the procedure (i.e. the state selection)
as done in \Eq{n-ave},
the detector density matrices $\rho_D^{(n)}$ and $\rho_D^{(n\pm 1)}$
in \Eq{n-ave} can be well approximated by $\rho_D^{(0)}$, i.e., the local thermal
equilibrium reservoir state.
Under this consideration, the detector fluctuation correlation functions become,
respectively,
$\la f^{\dg}(t)f(\tau)\ra = C^{(+)}(t-\tau)$,
$\la f(t)f^{\dg}(\tau)\ra = C^{(-)}(t-\tau)$,
and $\la F(t)F(\tau)\ra = C(t-\tau)=C^{(+)}(t-\tau)+C^{(-)}(t-\tau)$.
Here, $\la \cdots \ra$ stands for $\mb{Tr}_D [(\cdots)\rho_D^{(0)}]$.

Under the Markovian approximation, the time integral in \Eq{ME-2}
is replaced by $\frac{1}{2}\int^{\infty}_{-\infty}$.
Substituting \Eqs{L-H} and (\ref{n-ave}) into \Eq{ME-2}, we obtain
\bea\label{ME-3}
\dot{\rho}^{(n)}
   &=&  -i {\cal L}\rho^{(n)} - \frac{1}{2}
        \left\{  [Q\ti{Q}\rho^{(n)}+\mb{H.c.}] \right. \nl
   & &   - [\ti{Q}^{(-)}\rho^{(n-1)}Q+\mb{H.c.}]   \nl
   & &      \left.  - [\ti{Q}^{(+)}\rho^{(n+1)}Q+\mb{H.c.}] \right\} .
\eea
Here $\ti{Q}^{(\pm)}=\ti{C}^{(\pm)}({\cal L})Q$,
$\ti{C}^{(\pm)}({\cal L})=\int^{\infty}_{-\infty} dt C^{(\pm)}(t) e^{-i{\cal L}t}$,
and $\ti{Q}=\ti{Q}^{(+)}+\ti{Q}^{(-)}$.
Under the wide-band approximation for the detector reservoirs,
the spectral function $\ti{C}^{(\pm)}({\cal L})$ can be explicitly
carried out as \cite{Li04}:
$\ti{C}^{(\pm)}({\cal L})
  =  \eta \left[x/(1-e^{-x/T}) \right]_{x=-{\cal L}\mp V}$,
where $\eta=2\pi g_Lg_R$, and $T$ is the temperature.
In this work we will use the unit system of $\hbar=e=k_B=1$.
In \Eq{ME-3} the terms in $\{\cdots\}$ describe the fluctuation effect
of the forward and backward electron tunneling through the detector
on the measured system.
In particular,
the Liouvillian operator ``${\cal L}$" in $\ti{C}^{(\pm)}({\cal L})$
characterizes the energy transfer
between the detector and the measured system, which correlates the energy
(spontaneous) relaxation of the measured system
with the inelastic electron tunneling in the detector.
At high-voltage limit, formally $V\gg {\cal L}$, the spectral function
$\ti{C}^{(\pm)}({\cal L})\simeq \ti{C}^{(\pm)}(0)$, and \Eq{ME-3} reduces
to the previous result derived by Gurvitz {\it et al}
\cite{Gur97,Moz02,Gur03,Goa01a}.


In the following, we specify the measured system as a pair of coupled
quantum dots (a solid-state charge qubit), described by the Hamiltonian
$ H_{\rm qu} =  \epsilon_a |a\ra\la a| + \epsilon_b |b\ra\la b|
+ \Omega(|b\ra\la a|+|a\ra\la b|) $.
Introduce $\epsilon=(\epsilon_a-\epsilon_b)/2$,
and set $(\epsilon_a+\epsilon_b)/2$ as the reference energy.
The qubit eigen-energies are obtained as
$E_1=\sqrt{\epsilon^2+\Omega^2}\equiv \Delta/2$,
and $E_0=-\sqrt{\epsilon^2+\Omega^2}=-\Delta/2$.
Correspondingly, the eigenstates are
$|1\ra=\cos\frac{\theta}{2}|a\ra+\sin\frac{\theta}{2}|b\ra$
for the excited state,
and $|0\ra=\sin\frac{\theta}{2}|a\ra-\cos\frac{\theta}{2}|b\ra$
for the ground state, where $\theta$ is introduced by
$\cos\theta=2\epsilon/\Delta$, and $\sin\theta=2\Omega/\Delta$.
The coupling between the qubit and detector is characterized by
$H'=QF$, where $Q = {\cal T}+\chi |a\ra\la a|$,
and $F=\sum_{k,q} (c^{\dg}_kd_q + \mb{H.c.})$.

With the knowledge of $\rho^{(n)}(t)$, one is able to carry out the various
readout characteristics of the detector.
In the strong projective measurement regime (e.g. $\Omega=0$),
the measurement-induced wavefunction collapse of the qubit
can be perfectly manifested by the probability distribution function
$P(n,t)\equiv \mb{Tr}[\rho^{(n)}(t)]$.
Switching on $\Omega$ such that $1/\Omega$ is comparable to or smaller than
the measurement time \cite{Sch98}, the qubit state oscillation cannot be
read out by a series of single shot measurement.
In this regime, the continuous weak measurement is
an alternative approach to register the qubit oscillations.
In the remained part of the paper, we calculate the output current and noise
spectrum based on \Eq{ME-3}.

Straightforwardly, the average current flowing through the detector
can be generally expressed as
\bea\label{it}
I(t) =  \sum_n n
\mb{Tr}[\dot{\rho}^{(n)}(t)]
     = \frac{1}{2}\mb{Tr}[\bar{Q}\rho Q + \mb{H.c.} ] ,
\eea
where $\bar{Q}\equiv \ti{Q}^{(-)}-\ti{Q}^{(+)}$.
For symmetric qubit (i.e., $\epsilon=0$ or $\theta=\pi/2$),
the stationary current reads
\bea\label{Iss}
I_{s}=g_0V+g_1V\left[1-\frac{\Delta}{V}\frac{G^{(-)}}{G^{(+)}} \right] .
\eea
Here $g_0=\eta ({\cal T}+\chi/2)^2$, $g_1=\eta (\chi/2)^2$, and
$ G^{(\pm)} = \frac{1}{2}\left[F^{(+)}(\Delta,V)
                        \pm F^{(-)}(\Delta,V) \right] $,
with
$F^{(\pm)}(\Delta,V)\equiv (\Delta\pm V)\coth(\frac{\Delta\pm V}{2T})$.
We notice that \Eq{Iss} coincides with the result derived in
Ref.\ \onlinecite{Shn02}, but differs from that in Ref.\ \onlinecite{Sta03}.
The former was obtained on the basis of real-time Green's function
diagram technique,
while the latter was resulted from the quantum trajectory technique
under rotation-wave approximation.
In addition to the measurement current, in the following we detail
the studies of output noise spectrum in the regime of continuous
weak measurement, where a number of remarkable new features will be revealed.

\begin{figure}[h]
\begin{center}
\centerline{\includegraphics [scale=0.3,angle=0] {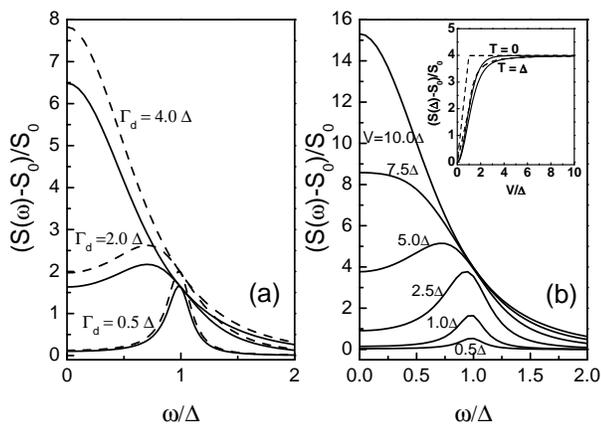}}
\end{center}
\caption{(a) Noise spectrum in the presence (solid curves) and
absence (dashed curves) of the qubit relaxation.
(b) Voltage effect on the noise spectrum, particularly on the
peak-to-pedestal ratio (inset, where the solid and dashed curves
correspond to the presence and absence of the qubit relaxation).
The results in (a) and (b) are obtained, respectively, by altering
$\chi$ (for a fixed voltage $V=2\Delta$ ) and the voltage $V$
(for a fixed $\chi=0.1\Delta$).
Other parameters are $g_L=g_R=2.5/\Delta$, and $T=\Delta$. }
\end{figure}

The noise spectrum can be calculated using the MacDonald's formula
\cite{Gur03}
\bea
S(\omega) = 2\omega \int^{\infty}_{0} dt \sin \omega t \frac{d}{dt}
            \left[\la n^2(t)\ra - (\bar{I} t)^2\right] ,
\eea
where $\bar{I}$ is the average current over time,
and $\la n^2(t)\ra = \sum_n n^2 P(n,t)$.
It can be shown that
\bea
\frac{d}{dt} \la n^2(t)\ra = \mb{Tr}\left[ \bar{Q} \hat{N}(t)Q
             + \frac{1}{2}\ti{Q}\rho(t)Q + \mb{H.c.}\right] ,
\eea
where $\hat{N}(t)\equiv \sum_n n \rho^{(n)}(t)$,
which can be calculated via its equation of motion
\bea
\frac{d\hat{N}}{dt} = -i {\cal L}\hat{N}
   -\frac{1}{2}\left[Q,\ti{Q}\hat{N}-\hat{N}\ti{Q}^{\dg} \right]
   + \frac{1}{2}(\bar{Q}\rho Q +\mb{H.c.}) .
\eea
For symmetric qubit, it would be desirable to carry out the explicit result.
Denoting $S(\omega)=S_0+S_1(\omega)+S_2(\omega)$,
the result reads
\begin{subequations}\label{SW}
\bea
S_0 &=& 2I_0 \coth\frac{V}{2T} + \frac{\chi^2 \eta}{2} \nl
          & &    \times \left[ G^{(+)}-\frac{\Delta^2}{G^{(+)}}
                 -V\coth\frac{V}{2T} \right] ,\\
S_1(\omega) &=& \left[1-\frac{\Delta}{2V}\frac{G^{(-)}}{G^{(+)}} \right]
       \frac{I^2_d\Gamma_d\Delta^2}{(\omega^2-\Delta^2)^2
       +\Gamma_d^2\omega^2} ,\\
S_2(\omega) &=& \chi^2 \eta \left[\Gamma_d D_z
             +\gamma \bar{I} \right]
                \frac{G^{(-)}}{\omega^2+\Gamma_d^2}  .
\eea
\end{subequations}
Here three currents are defined as $I_0=(I_a+I_b)/2$, $I_d=I_a-I_b$,
and $\bar{I}= I_0-\frac{1}{4} \eta\chi^2\Delta G^{(-)}/G^{(+)}$,
with $I_a=\eta ({\cal T}+\chi)^2V$ and $I_b=\eta {\cal T}^2V$ being
the detector currents corresponding to
qubit states $|a\ra$ and $|b\ra$, respectively.
Other quantities in \Eq{SW} are introduced as:
$\Gamma_d=\frac{\eta\chi^2}{2}G^{(+)}$,
$\gamma=\frac{\eta\chi^2}{2}\Delta$
and $D_z= -\Delta\sqrt{I_aI_b}/G^{(+)}-\eta\chi^2 G^{(-)}/4$.
The three noise spectrum components are, respectively,
(i) the zero-frequency noise $S_0$,
(ii) the Lorentzian spectral function $S_1(\omega)$
with a peak around the qubit Rabi frequency $\omega=\Delta$,
and (iii) $S_2(\omega)$ completely originating from the qubit
relaxation induced inelastic tunnelling effect in the detector.
In addition to $S_2(\omega)$, the qubit relaxation also manifests
its effect in $S_0$ and $S_1(\omega)$,
i.e., giving rise to the second term of $S_0$
and reducing the pre-factor in $S_1(\omega)$ from unity.
If the qubit relaxation induced inelastic effect is neglected,
or at the limit of high bias voltage $V\gg\Delta$,
\Eq{SW} returns to the known result of previous work
\cite{Kor01a,Goa01a}.

The measurement-induced relaxation effects of the qubit are further shown
in Fig.\ 1. The major effect of the qubit relaxation shown in Fig.\ 1(a)
is lowering the entire noise spectrum, in qualitative consistence
with the finding by Gurvitz {\it et al} \cite{Gur03}, where an
{\it external thermal bath} is introduced to cause qubit relaxation.
However, the {\it spontaneous} relaxation discussed here
does not diminish the telegraph noise peak near zero frequency
in the incoherent case,
which implies the surviving of the Zeno effect,
in contrast to the major conclusion of Ref.\ \onlinecite{Gur03}.
Also, the transition behavior from the coherent to the incoherent regime
is different.
Figure 1(b) shows the voltage effect that
the coherent peak around $\omega=\Delta$ reduces as
decreases the measurement voltage.
Interestingly, this effect alters the fundamental upper bound limit of 4
for the value of the peak-to-pedestal (``signal-to-noise") ratio,
$[S(\Delta)-S_0]/S_0$,
which was found by Korotkov {\it et al.} at the high voltage limit
(see the inset) \cite{Kor01a}.

\begin{figure}[h]
\begin{center}
\centerline{\includegraphics [scale=0.5,angle=0] {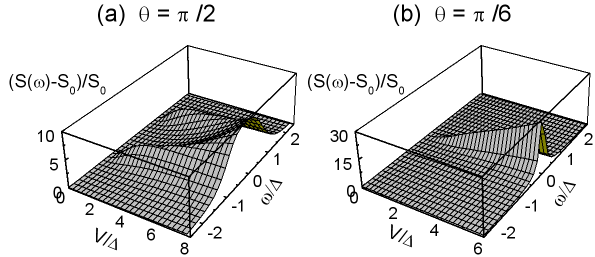}}
\end{center}
\caption{ 3D-plot of the noise spectra for (a) the symmetric qubit,
and (b) the asymmetric qubit.
The adopted parameters are $g_L=g_R=2.5/\Delta$, $\chi=0.1\Delta$,
and $T=\Delta$. }
\end{figure}

The voltage effect is further shown in Fig.\ 2 by the
3D-plot of the scaled spectra for different qubit symmetries.
In contrast to the present result,
we notice that in Ref.\ \onlinecite{Sta03} no spectral structure was found,
i.e., $S(\omega)-S(\infty)=0$, in the wide range of $V<10\Delta$
for the symmetric qubit ($\theta=\pi/2$).
However, Shnirman {\it et al} showed the existence of the coherent peaks
at $\omega=\pm\Delta$ for voltage higher than $\Delta$
\cite{Shn02}.
For asymmetric qubit as shown in Fig.\ 2(b), the coherent peaks at
$\omega=\pm\Delta$ are destroyed and a peak around $\omega=0$ is formed.
This transition originates from the breakdown of the resonant condition,
which replaces the Rabi oscillation of the qubit by incoherent jumping.

Finally, simple analysis in limiting cases can provide
additional insight into the correlation between the detector and the qubit.
At zero temperature, we obtain a pre-factor 1/2 in $S_1(\omega)$ in
the low-voltage regime ($V<\Delta$).
This result is in sharp difference from previous conclusions:
In Ref.\ \onlinecite{Shn02} such kind of contribution vanishes,
while in Ref.\ \onlinecite{Sta03} it
does not exist at all in a much wider range of voltage.
At the same limit, \Eq{SW} also predicts non-vanishing $S_2(\omega)$
and non-zero correction to the Schottky shot noise $2I_0$ in $S_0$.
Remarkably, all these contributions were in absence from the previous work
\cite{Shn02,Sta03}, and the reason was attributed to the complete relaxation
of the measured qubit to its ground state.
Here we understand our distinct result as follows.
Consider the key quantity $\la n^2(t)\ra=\sum_n n^2 P(n,t)$.
Despite the fact that the qubit would relax to its ground state
under the concerned limit, the fluctuation of $\la n^2(t)\ra$ remains
according to \Eq{ME-3}, since the {\it conditional} qubit state
$\rho^{(n)}(t)$ is {\it not} at all stationary.
In particular, the non-zero off-diagonal elements of $\rho^{(n)}(t)$
contain the information of qubit coherence, which gives rise to
the peak structure of the noise spectrum.
Therefore, differing from the previous work \cite{Shn02,Sta03}
and even going beyond the very recent debate \cite{Ave04},
we conclude here that at zero temperature and even in low bias voltage,
the detector output noise spectrum still contains {\it excess} components
in addition to the Schottky shot noise, due to the quantum fluctuations
induced by the coupling of the detector and the qubit.
This novel feature may deserve further confirmation in future work.

In summery, we have generalized the validity range of
the quantum measurement theory
developed by Gurvitz {\it et al.} to arbitrary voltage and temperatures.
The generalized theory properly accounts for the energy transfer between
the detector and the measured system.
Its application to charge qubit measurement reveals
a number of interesting new readout characteristics associated with the
new treatment of the correlation between the detector and the qubit.


\vspace{5ex}
{\it Acknowledgments.}
Support from the National Natural Science Foundation of China,
the Major State Basic Research Project No.\ G001CB3095 of China,
and the Research Grants Council of the Hong Kong Government
are gratefully acknowledged.


\end{document}